\documentclass[12pt]{article}
\input epsf
\begin{document}
\baselineskip16pt
\thispagestyle{empty}
\date{}
\title{Isotropization of Bianchi type models \\
and a new FRW solution in Brans-Dicke theory}
\author{Jorge L. Cervantes-Cota\thanks{e-mail:jorge@nuclear.inin.mx} \, 
and Marcos Nahmad \\
Departamento de F\'{\i}sica, \\
Instituto Nacional de Investigaciones Nucleares (ININ) \\
P.O. Box 18-1027, M\'exico D.F. 11801, M\'exico} 
\maketitle
\begin{abstract}
Using scaled variables we are able to integrate an equation valid for 
isotropic and anisotropic Bianchi type I, V, IX models in 
Brans-Dicke (BD) theory.  We analyze known and new solutions for these 
models in relation with the possibility that anisotropic models 
asymptotically isotropize, and/or possess inflationary 
properties. In particular, a new solution of curve ($k\neq0$) 
Friedmann-Robertson-Walker (FRW) cosmologies in Brans-Dicke theory 
is analyzed. 

\end{abstract}


\section{INTRODUCTION \label{intro}} 

The universe is nowadays at big scales homogeneous and isotropic as measured in 
the CMBR by the COBE satellite \cite{Cobe94}, and must also has been having 
these properties since, at least, the era of nucleosynthesis \cite{Isonu76}. In 
order to explain the isotropy of the universe from theoretical anisotropic 
models, many authors have considered the Bianchi models that can in principle 
evolve to a Friedmann-Robertson-Walker (FRW) cosmology.  It has been shown that 
some Bianchi models in General Relativity (GR) tend to their isotropic  
solutions, up some extent \cite{CoHa73,BaSo86}, and even that they can explain 
the level of anisotropy measured by COBE \cite{Ba95}.  Motived by these facts, 
we have been working in Brans-Dicke (BD) theory \cite{BrDi61} to 
investigate if Bianchi universes 
are able to isotropize as the universe evolves, and if its evolution can 
be inflationary.  In previous investigations 
we have shown that anisotropic, Bianchi type I, V, and IX models tend to 
isotropize as models evolve \cite{ChCe95,MiWa95}.  However, this may happen 
for some restrictive values of $\omega$ in the cases of
Bianchi type I and IX, and only Bianchi type V model 
can accomplish an isotropization mechanism within BD current constraints 
\cite{CeCh99} on $\omega$.  It has been also shown that the isotropization 
mechanism in the Bianchi type V model can be inflationary, without the presence 
of any cosmological constant, when small values for the coupling constant 
$\omega$ are considered \cite{CeCh99}, as in the case of some 
induced gravity (IG) models \cite{ig2,ig3}.  We have recently 
shown, however, that the isotropization mechanism can be attained with 
sufficient amount of e-folds of inflation, only for negative values of 
$\omega$.  In the present report we review and generalize some of the main 
results and present a new $k\neq0$ FRW solution in BD theory.  

This paper is organized as follows.  In section \ref{bdsol} the BD Bianchi
type I, V and IX equations are presented.   In section \ref{anisol} we review 
the main results on these models and present a new solution to curve $k\neq0$ 
FRW cosmologies.  Finally, conclusions are in section \ref{con}.

\section{ANISOTROPIC EQUATIONS FOR BIANCHI MODELS \label{bdsol}}

In previous investigations \cite{ChCe95,CeCh99,Ce99} we have used scaled 
variables, in terms of which our solutions have been given, therefore  
following we use them: the scaled field 
$\psi \equiv \phi a^{3(1-\nu)}$, a new cosmic time parameter 
$d\eta \equiv a^{-3\nu} dt$, $()^\prime\equiv \frac{d}{d\eta}$, 
the `volume' $a^{3}\equiv a_{1}a_{2}a_{3}$, and the Hubble parameters 
$H_{i}\equiv {a_{i}}^\prime /a_{i}$ corresponding to the scale factors 
$a_{i}=a_{i}(\eta)$ for $i=1,2,3$.  We assume comoving coordinates and 
a perfect fluid with barotropic equation of state, $p=\nu \rho$, 
where $\nu$ is a constant. Using these definitions one obtains the 
cosmological equations for Bianchi type I, V and IX 
models (in units with $G = c = 1$):
\begin{equation}
\label{a123}
(\psi H_i)^\prime -  \psi a^{6 \nu} R_{ij}  \, = \, 
\frac{8 \pi}{3+2\omega}  [1 + (1 - \nu) \omega] \, \rho  a^{3(1+\nu)}
 ~~~ {\rm for} ~~~ i=1,2,3. 
\end{equation}

\begin{eqnarray}
\label{h123} 
&&H_{1}H_{2} + H_{1}H_{3} + H_{2}H_{3} + 
[1+(1-\nu)\omega] \, \left(H_{1}+H_{2}+H_{3}\right) 
\frac{\psi^\prime}{\psi}
\nonumber \\[2pt]
&&- (1-\nu)[1+\omega(1-\nu)/2] (H_{1}+H_{2}+H_{3})^{2} 
-\frac{\omega}{2} \left( \frac{\psi^\prime}{\psi}\right)^{2} 
-\frac{R_{j}}{2} a^{6 \nu}  \nonumber \\[2pt]
&& = \,\ 8 \pi \, \frac{\rho a^{3(1+\nu)}}{\psi} \,  \,\ ,
\end{eqnarray}

\begin{equation}
\label{psi} 
\psi^{\prime \prime} + (\nu-1) a^{6 \nu} R_{j} \, \psi =  
\frac{8 \pi}{3+2\omega}  [2(2-3\nu)+3(1-\nu)^{2} \omega ] \, 
\rho  a^{3(1+\nu)} \,\ ,
\end{equation}
where a column sum is given by $R_{j}\equiv \Sigma_{i} R_{ij}$, where  
$j=$I, V or IX and  
\begin{equation}
\label{curv}
\matrix {
& &~{}^I &~~~~~~~~~{}^V~~~~~~~~ &~{}^{IX} \cr 
& & & & \cr
& &0 & 2  / a_1^2 &
[a_1^4-a_2^4-a_3^4 + 2 a_2^2 a_3^2] / (-2 a^6) \cr
& & & & \cr
& R_{ij}~ = &0 & 2  / a_1^2 & [a_2^4-a_3^4-a_1^4 + 2 a_1^2 a_3^2] 
/ (-2 a^6) \cr
& & & & \cr
& &0 & 2  / a_1^2 & [a_3^4-a_1^4-a_2^4 + 2 a_1^2 a_2^2] / (-2 a^6) \cr} 
\end{equation}
For the Bianchi type V model one has the additional constraint:    
$H_{2} + H_{3} \,\ = \,\ 2 H_{1} $, implying that $a_{2}$ and $a_{3}$ are 
inverse proportional functions, $a_{2}a_{3}= a_{1}^{2}$; note that 
the mean Hubble parameter, $H\equiv \frac{1}{3}(H_{1}+H_{2}+H_{3})$, is 
for this Bianchi type model $H=H_{1}$. 

Additionally, the continuity equation yields:
$\rho a^{3(1+\nu)}={\rm const.}\equiv M_{\nu}$, $M_{\nu}$ being a 
dimensional constant depending on the fluid present. The vacuum case is 
attained when $M_{\nu}=0$.
  
The system of ordinary differential equations, Eqs.  
(\ref{a123}-\ref{psi}), can be once integrated to get\footnote{A similar
equation, that is valid only for the Bianchi type V model, was derived 
in Ref. \cite{Ce99}.  Now we generalize that result in such a way that 
Eq. (\ref{psi1}) is valid for Bianchi models I, V and IX, as well for FRW 
cosmologies.}:
\begin{eqnarray}
\label{psi1}
&&\psi \, \psi^{\prime \prime} -  \frac{2}{3(1-\nu)} {\psi^{\prime}}^{2} -
\frac{2(1-3\nu)}{3(1-\nu)} \, [m_{\nu} (1-3\nu) \eta + \delta ] \, 
\psi^{\prime} + [2+(1-\nu)(1+3\nu)\omega]m_{\nu} \, \psi + \nonumber \\
&&\frac{2}{3(1-\nu)} \, [2-3\nu + \frac{3}{2} (1-\nu)^{2} \omega] \, 
[m_{\nu} (1-3\nu) \, \eta + \delta ]^{2} + (1-\nu) (h_{1}^{2}+ h_{2}^{2}+
h_{3}^{2}) = 0 \,\ ,
\end{eqnarray}
where $\delta $ is an  integration constant,  
$m_{\nu}\equiv \frac{8 \pi M_{\nu}}{3+2 \omega}$, and  the Hubble rates are 
written as follows (similar to the Bianchi type I model deduced in 
Ref. \cite{RuFi75}):
\begin{equation} 
\label{hi}
H_i \, = \, {1 \over 3} \left( H_{1}+H_{2}+H_{3} \right) + 
{h_i \over \psi } = 
\frac{\psi^{\prime} - (1-3\nu) m_{\nu} \, \eta -  \delta  + 
3 (1-\nu) h_i}{3 (1-\nu) \psi} \, ,
\end{equation}
where the $h_i$'s are some unknown functions of $\eta$ that determine the 
anisotropic character of the solutions.  If $h_i=0$ for $i=1, 2$, $3$ 
simultaneously, no anisotropy is present, which is the case of FRW 
cosmologies. Furthermore, Bianchi models obey the condition
\begin{equation}
\label{resth123}
 h_1 + h_2 + h_3 = 0   
 \end{equation}
to demand consistency with Eq. (\ref{hi}).  For the Bianchi type V model 
one has additionally that $h_1 = 0$, since $H_{1} = H$ as mentioned above.  

Equations (\ref{hi}) and (\ref{resth123}) imply that the mean Hubble 
parameter is determined by $\psi$ alone:
\begin{equation}
\label{psih}
3 H \, = \, H_{1} + H_{2} + H_{3} \, = \, \frac{1}{1-\nu}  \,  
\left[ \frac{\psi^{\prime}}{\psi}  - 
\frac{(1-3\nu) m_{\nu} \, \eta + \delta }{\psi} \right]   \, .
\end{equation}

In order to analyze the anisotropic character of the solutions, we  
consider the anisotropic shear, 
$\sigma \equiv - (H_1 - H_2)^2 - (H_2 - H_3)^2 - (H_3 - H_1)^2$. $\sigma = 0$ 
is a necessary condition to obtain a FRW cosmology since it implies 
$H_1 = H_2 = H_3$, cf. Ref. \cite{ChCe95,ChCeNu91}.  If the sum of 
the squared differences of the Hubble expansion rates 
tends to zero, it would mean that the anisotropic scale factors tend to 
a single function of time which is, certainly, the scale factor of  
a FRW solution.  

The anisotropic shear becomes, using 
Eqs. (\ref{hi}) and (\ref{resth123}),
$ \sigma ( \eta)  \,\ = - \frac{3 (h_{1}^{2}+ h_{2}^{2}+ h_{3}^{2})}{\psi^2}$, 
or the dimensionless shear parameter \cite{WaEl97}, using Eq. (\ref{psih}):
\begin{equation}
\label{dlsig}
\frac{\sigma}{H^{2}} = - \frac{27 (1-\nu)^{2} (h_{1}^{2}+ h_{2}^{2}+ 
h_{3}^{2})}{[\psi^{\prime}-(1-3\nu) m_{\nu} \eta -\delta ]^{2}} \,\  .
\end{equation}
If the above equations admit solutions such that   
$\sigma/H^{2} \to 0$ as $\eta \to \infty$  ($t \to \infty$), then  
one has time asymptotic isotropization solutions, similar 
to solutions found for the Bianchi models in GR \cite{BaSo86}.  In fact, 
one does not need to impose an asymptotic, infinity condition, but just 
that $\eta \gg \eta_{*}$, where $\eta_{*}$ is yet some arbitrary value 
to warrant that $\sigma/H^{2}$ can be bounded from above.  

\section{ANISOTROPIC AND ISOTROPIC SOLUTIONS \label{anisol}}

The problem to find solutions of Bianchi and FRW models in the BD theory has 
been reduced to solve the coupled system of equations (\ref{psi1}), (\ref{hi})
and (\ref{a123}).  Let us present in the following subsections the known and 
new solutions. 
 
\subsection{Bianchi type I model}

For this Bianchi model the known, the general solution is found in which the 
$h_i$'s are constants, then Eq. (\ref{psi1}) is decoupled from Eqs. (\ref{a123})
and (\ref{hi}), and the solution is 
$\psi \,= \, A_{I} \, \eta^2 + B_{I} \, \eta + C_{I}$, where the constants
$A_{I}, B_{I},  C_{I}$ are reported elsewhere \cite{ChCe95,RuFi75}.  This 
model can be solved in a general way since the curvature sum column 
$R_I=0$, then Eq. (\ref{psi}) can be directly integrated. 
This solution represents a particular solution of  Eq. (\ref{psi1}).   
Direct substitution of $\psi$ into Eq. (\ref{hi}) gives the Hubble 
rates, and into Eq. (\ref{dlsig}) shows that solutions isotropize as time 
evolves, that is, $\sigma/H^{2} \to 0$ as $\eta \to \infty$, see 
Ref. \cite{ChCe95}.  However, the isotropization mechanism is only possible
for solutions such that $\Delta_{I}\equiv B_{I}^{2}-4 A_{I} C_{I}$
\begin{eqnarray}
\label{discri1}
\Delta_{I} = \frac{2(2-3\nu)+3(1-\nu)^{2}\omega}{3(1-\nu)^{2}(3+2\omega)}
&& [ \frac{(1-\nu)^{2}B_{I}^{2}}{2(2-3\nu)+3(1-\nu)^{2}\omega}
-2(1-3\nu) \delta \, B \nonumber \\ 
&&+ [2(2-3\nu)+3(1-\nu)^{2}\omega] \delta^{2} \nonumber \\ 
&&+ 3(1-\nu)^{2} (h_{1}^{2}+h_{2}^{2}+h_{3}^{2}) 
]
\end{eqnarray}
is negative \cite{CeCh99}.  Then,
some restrictions on $\omega$ apply. For instance, in Dehnen's IG 
theory \cite{ig3} $\omega \ll 1$, then the isotropization 
mechanism is not possible in this Bianchi model \cite{CeCh99}.

\subsection{Bianchi type V model}

This Bianchi model is more complicated because curvature terms are 
different than zero.  Still, it is possible to find particular solutions of   
Eq. (\ref{psi1}), since for this model the $h_i$'s are constants too, and 
this equation is decoupled as is the case of Bianchi type I 
model.  The general solution of this model should be obtained through 
the general solution of Eq. (\ref{psi1}), yet unknown.  We have found 
a particular solution that is again of the form 
$\psi \,= \, A_{V} \, \eta^2 + B_{V} \, \eta + C_{V}$, where the 
constants $A_{V}, B_{V},  C_{V}$ are reported in 
Refs. \cite{ChCe95,Ce99}.  This solution is such that 
\begin{equation}
\label{discri5}
\Delta_{V} = {{-8 (1-3\nu)^2} \over 
{18 \nu+(1+3\nu)^2 \omega}}{h_2^2}   
\end{equation}  
is negative for $\omega>-18\nu/(1+3\nu)^{2}$, so the solution tends to 
isotropic solution within BD theory constraints \cite{Wi93}, $\omega\ge 500$, 
that is $\sigma/H^{2} \to 0$ as $\eta \to \infty$.  For this
Bianchi type model, Dehnen's IG theory \cite{ig3} can achieve
an isotropization mechanism \cite{CeCh99}.  An inflationary behavior may be   
observed in type V models, but to get enough e-foldings of 
inflation ($N \sim 68$) one must demand that $\omega \le -\frac{3}{2}$ 
\cite{Ce99} in consistency with previous results \cite{LeFr94}.

\subsection{Bianchi type IX model}

Bianchi type IX model is the most complicated to solve, since curvature
terms involve quartic polynomials of the scale factors, see 
Eq. (\ref{curv}).  For this Bianchi model it 
implies, by imposing the condition that $h_i$'s are 
constants, severe algebraic constraints on the scale factors, so it seems more 
likely that $h_i$'s are functions.  This explains why no totally anisotropic 
($H_{1}\neq H_{2}\neq H_{3}$) solution
has been found yet.  In Ref. \cite{ChCe95} we have analyzed the case when 
the polynomial solution for $\psi$ is valid.  In this case, unfortunately,
we could not found explicitly the values of the constants 
$A_{IX}, B_{IX}, C_{IX}$.  If this solution
is valid, however, one has that 
$h_{1}^{2}+ h_{2}^{2}+ h_{3}^{2} = D \eta^2 + F \eta + G$, where
$D, F, G$ are constants.  Accordingly, Eq. (\ref{dlsig}) 
indicates that the solution must tend, 
as time evolves, to the positive curvature FRW solution, i.e. one has 
again that $\sigma/H^{2} \to 0$ as $\eta \to \infty$.  However, a 
definitive answer will arrive by obtaining explicitly the values 
$A_{IX}, B_{IX}, C_{IX}$.

\bigskip

The only possible solutions for Bianchi type I and V models imply 
that the $h_i$'s are constants, whereas for type IX they are unknown 
functions of $\eta$.  An explanation of this fact resides in the property  
that Bianchi type I and V models have curvature terms of FRW 
type, whereas type IX has a very much complicated 
form,  see Eq. (\ref{curv}).  So the things, it seems that the most 
general solution of Eq. (\ref{psi1}) with $h_i$'s constants 
would give general solutions for Bianchi models I and V.  The particular 
quadratic-polynomial solution of Eq. (\ref{psi1}) represents in the case of 
Bianchi type I model its the general solution, whereas for Bianchi model V 
it is only a particular solution.  Then, other particular 
solutions, possibly of non-polynomial nature, are expected to be found for 
Eq. (\ref{psi1}) that will reveal new aspects of Bianchi type V 
model.  Finally, for Bianchi type IX model the quadratic-polynomial can be a 
possible solution, not yet confirmed. However, solutions with 
$h_i(\eta)$ valid for the Bianchi type IX model are almost impossible to 
find because of the complexities involved in the curvature terms. 

\subsection{FRW solutions}

It turns out that Eq. (\ref{psi1}) is also valid for the FRW models when the
anisotropic parameters vanish, i.e. $h_{1} = h_{2}= h_{3} = 0$.  Solutions of 
this equation solve FRW cosmologies in BD theory 
\cite{ChCe95,bdfrwo}.  The known solutions for $\psi$ are quadratic 
polynomials in $\eta$ as well.  The general 
flat ($k=0$) solution is a particular solution of Eq. (\ref{psi1}) in which 
the coefficient of the quadratic polynomial term, $A$, is equal to the 
corresponding coefficient ($A_{I}$) of the Bianchi I case.  For 
curved ($k \neq 0$) FRW cosmologies the known particular solution is such 
that the coefficient of $A$ of the quadratic polynomial term is equal to 
the corresponding coefficient ($A_{V}$) in the Bianchi V case.  In this
way, one can see a correspondence between anisotropic and isotropic solutions.

We have found a new solution of Eq. (\ref{psi1}) that is valid for 
$k \neq 0$ FRW cosmologies.  The new solution is:
\begin{equation}
\label{psifrw}
\psi = m_{\nu} \, \left(\frac{1-3\nu}{1+3\nu}\right)^{2} \left[ \kappa_{1} +
(1-3\nu)\, \eta \right]  \left[\kappa_{1} + 2 \eta + \kappa_{2} \, 
\left[\kappa_{1} + (1-3\nu) \, \eta \right]^{\frac{2}{1-3\nu}} \right]  
\,\  ,
\end{equation}
where $\kappa_{1}, \kappa_{2}$ are arbitrary integration constants. This
is the general solution of curved FRW cosmologies when the 
following relationships are valid:
\begin{eqnarray}
\label{wnu}
\delta  & = & 0  \nonumber \\[2pt]
\omega & = &\frac{- 18 \, \nu}{(1+3\nu)^2} 
\,\  .
\end{eqnarray}
Though the latter relationship constrains the range of possible values of 
$\omega$ and $\nu$, one can find values of physical interest, 
e.g. $\omega=-1$ that has some interest in string effective 
theories, see for instance Ref. \cite{Di98}. Moreover,  when 
$\nu \rightarrow -1/3$ one obtains the GR limit 
$\omega \rightarrow \infty$. Finally, one gets the dust 
model ($\nu_{{\rm dust}}=0$) in the limit when $\omega \rightarrow 0$, 
like in Dehnen's IG theory \cite{ig3}. 

The Hubble parameter is given by:   
\begin{equation}
\label{hfrw}
H= \frac{\kappa_{1} + \frac{1-9\nu}{1-3\nu}\, \eta + 
\kappa_{2} \, \left[\kappa_{1} + (1-3\nu) \, \eta \right]^{\frac{2}{1-3\nu}}}
{\left[ \kappa_{1} +
(1-3\nu)\, \eta \right]  \left[\kappa_{1} + 2 \eta + \kappa_{2} \, 
\left[\kappa_{1} + (1-3\nu) \, \eta \right]^{\frac{2}{1-3\nu}}\right]}
\,\  ,
\end{equation}
{}from which one can find the scale factor:
\begin{equation}
\label{afrw}
a= \left[\frac{-k}{\kappa_{2}}\right]^{\frac{1}{2(1-3\nu)}} 
\frac{\left[\kappa_{1} + 2 \eta + \kappa_{2} \, 
\left[\kappa_{1} + (1-3\nu) \, \eta \right]^{\frac{2}{1-3\nu}}
\right]^{\frac{1}{2(1-3\nu)}}}
{\left[\kappa_{1} + (1-3\nu) \, \eta \right]^{\frac{3 \nu}{(1-3\nu)^{2}}}}
\,\  .
\end{equation}

The BD field ($\phi = \psi a^{-3 (1-\nu)}$) is obtained through 
Eqs. (\ref{psifrw}) and (\ref{afrw}):
\begin{equation}
\label{phifrw}
\phi =  m_{\nu} \, \left(\frac{1-3\nu}{1+3\nu}\right)^{2}
\left(\frac{-\kappa_2}{k}\right)^{\frac{3(1-\nu)}{2(1-3\nu)}}
\frac{\left[\kappa_{1} + (1-3\nu) \, \eta 
\right]^{\frac{1+3\nu}{(1-3\nu)^{2}}} }
{\left[\kappa_{1} + 2 \eta + \kappa_{2} \, 
\left[\kappa_{1} + (1-3\nu) \, \eta \right]^{\frac{2}{1-3\nu}}
\right]^{\frac{1+3\nu}{2(1-3\nu)}}}
\,\  .
\end{equation}

Eqs. (\ref{afrw}) and  (\ref{phifrw}) imply that the sign of $\kappa_2$ is 
equal to the sign of $-k$ for most values of $\nu$.   For open ($k=-1$) models 
this implies that $\kappa_2 $ must be positive, and for closed ($k=+1$ ) 
models $\kappa_2$ must be negative which allows the solutions to 
(re)collapse: The value of $\kappa_2 $  determines the time of maximum
expansion, so it is very related to the mass ($m_\nu$) of the model.  On the
other hand,  $\kappa_1 $ represents a $\eta$-time shift.

\begin{figure}
\unitlength 1in
\centering
\begin{picture}(+6.5,6.5)(0,1)
\epsfxsize=5.0in \epsfbox{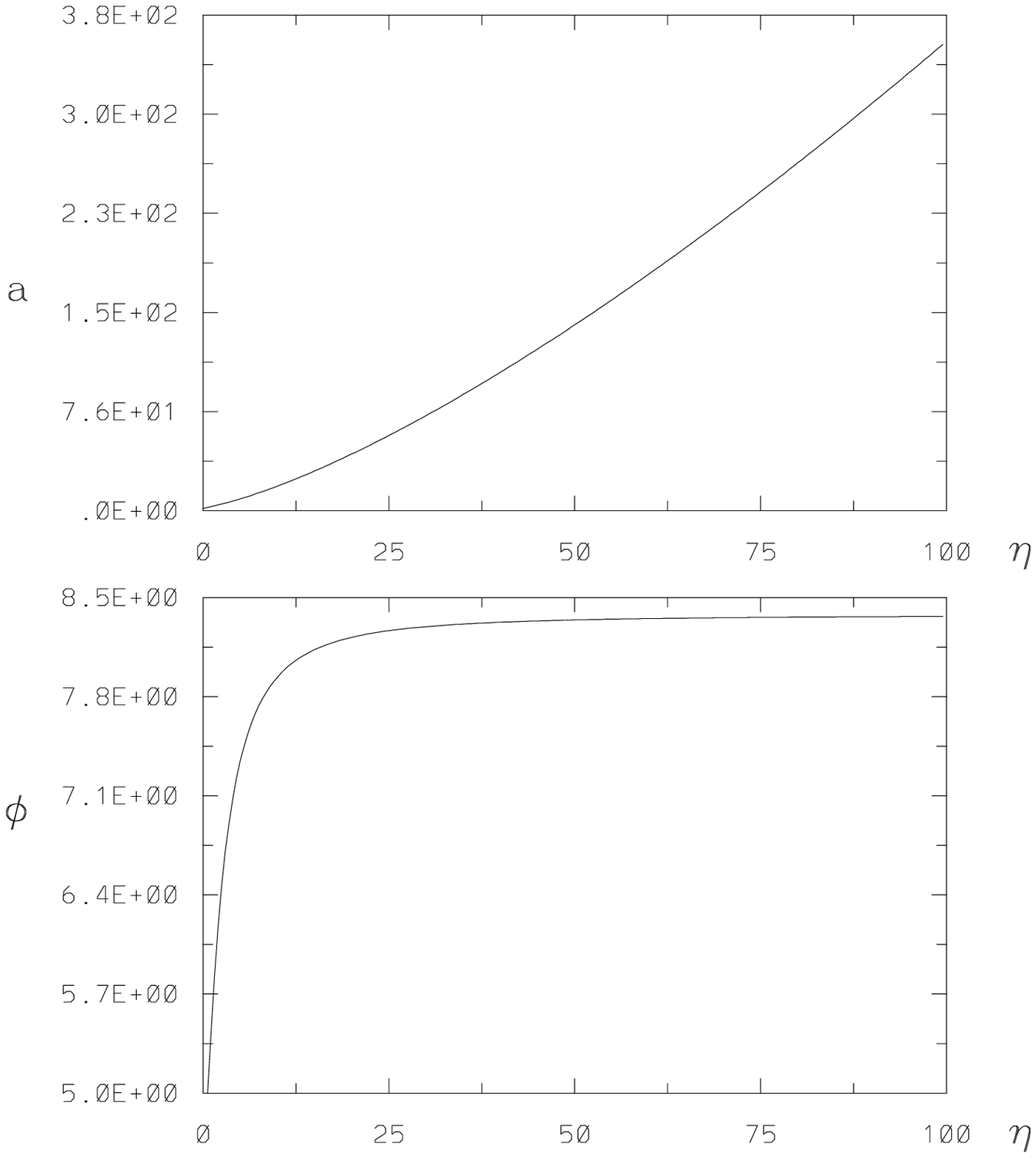}
\end{picture}
\vskip 1.5cm
\caption{The scale factor and BD field a function of time $\eta$.  The plots 
represent an open model with $\kappa_1=\kappa_2= -k =1$, and $\nu=\nu_1$ 
that implies $\omega=-1$.
\label{fig1} }
\end{figure}

\begin{figure}
\unitlength 1in
\begin{picture}(6.5,6.5)(0,1)
\epsfxsize=5in \epsfbox{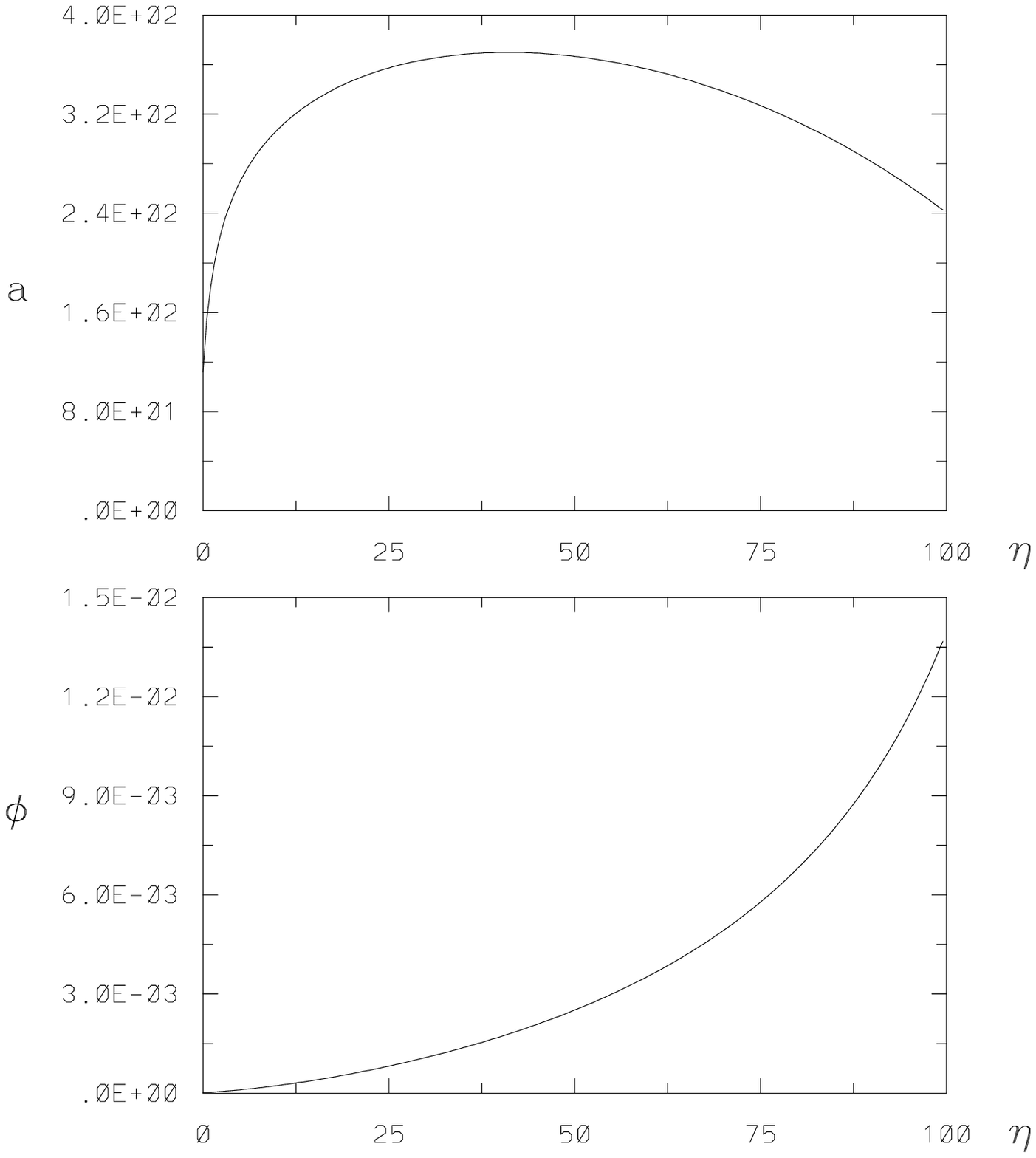}
\end{picture}
\vskip 1.5cm
\caption{The scale factor and BD field a function of time $\eta$.  The plots 
represent a closed model with $\kappa_1 = k =1$, $\kappa_2 = -0.001$, and 
$\nu=\nu_1$ that implies $\omega=-1$.  The model eventually recollapses.
\label{fig2}}
\end{figure}

Because of the mathematical form of Eq. (\ref{hfrw}) it is not possible
to have an inflationary era that lasts for a sufficient time period to
solve the horizon and flatness problems.  To show the behavior of the models 
we have plotted the above-given formulae for different values of $\omega$,
the curvature constant $k$, and integration constants $\kappa_1$ and 
$\kappa_2$.  We have chosen in all figures that 
$\phi^{\prime}\mid_{\eta=0}\equiv\phi^{\prime}_{0}= 0$ as initial 
condition.  Given a specific value for $\omega$ implies two possible
values of $\nu$ consistent with Eq. (\ref{wnu}).   For  $\omega=-1$, a 
value that makes BD theory to resemble string effective theories \cite{Di98},
it corresponds $\nu_1=(2-\sqrt{3})/3\approx 0.0893$ (that represents
a quasi dust model) and 
$\nu_2=(2+\sqrt{3})/3\approx 1.2440$.   In figures \ref{fig1} and 
\ref{fig2} we plotted the scale factor and BD field for $\nu_1$.  Figure 
\ref{fig1} shows an open model with $\kappa_1=\kappa_2= -k =1$, whereas 
figure \ref{fig2} is a closed model 
with $\kappa_1 = k =1$ and $\kappa_2 = -0.001$.  In closed models, the smaller 
$\kappa_2$ is, the later in time they will recollapse.  However, for models  
with $2/(1-3\nu) < 1$ one can find an upper limit on $\kappa_2$ such that 
models never recollapse (this happens, for instance, when $\nu=-0.3719$,  
making $\omega=500$, and if $\kappa_2 < -1$); the effect of negative pressure
avoids recollapse.

One can compute the asymptotic limit for $k=-1$ models, when 
$\eta \rightarrow \infty$.  There are two limit cases, when the quantity 
$2/(1-3\nu)$ is greater or smaller than 1, cf. terms in the 
equations above. In the former case one has that
\begin{eqnarray}
\label{limg}
a_{{\rm asimp1}} &=& \left(-k\right)^{\frac{1}{2(1-3\nu)}} 
\left[ (1-3\nu) \, \eta\right]^{\frac{1}{1-3\nu}} \nonumber \\[2pt]
\phi_{{\rm asimp1}} &=&  m_{\nu} \, \kappa_2 \,
\left(\frac{1-3\nu}{1+3\nu}\right)^{2}
 \left(-k \right)^{- \frac{3(1-\nu)}{2(1-3\nu)}} 
\, = \, {\rm const.}   \,\  ,
\end{eqnarray}
whereas in the latter case one gets
\begin{eqnarray}
\label{lims}
a_{{\rm asimp2}} &=& \left[\frac{-2k}{\kappa_{2}}\right]^{\frac{1}{2(1-3\nu)}} 
\left(1-3\nu \right)^{\frac{-3 \nu}{(1-3\nu)^{2}}}  \,
\eta^{\frac{1-9\nu}{2(1-3\nu)^{2}}}
\nonumber \\[2pt]
\phi_{{\rm asimp2}} &=&  2 \, m_{\nu} \, \frac{(1-3\nu)^{3 + 
\frac{9\nu(1-\nu)}{(1-3\nu)^{2}}}}{(1+3\nu)^{2}}
\left(\frac{-\kappa_2}{2 k}\right)^{\frac{3(1-\nu)}{2(1-3\nu)}} \, 
\eta^{\frac{(1+3\nu)^{2}}{2(1-3\nu)^{2}}}
\,\  .
\end{eqnarray}
The plots of figure \ref{fig1} tends to Eqs. (\ref{limg}); one observes that
$\phi$ tends to a constant value that can be fix to be $G^{-1}$ through the 
right choice of the constants $m_{\nu}$ and $\kappa_2$.

\begin{figure}
\unitlength 1in
\centering
\begin{picture}(+6.5,6.5)(0,1)
\epsfxsize=5.0in \epsfbox{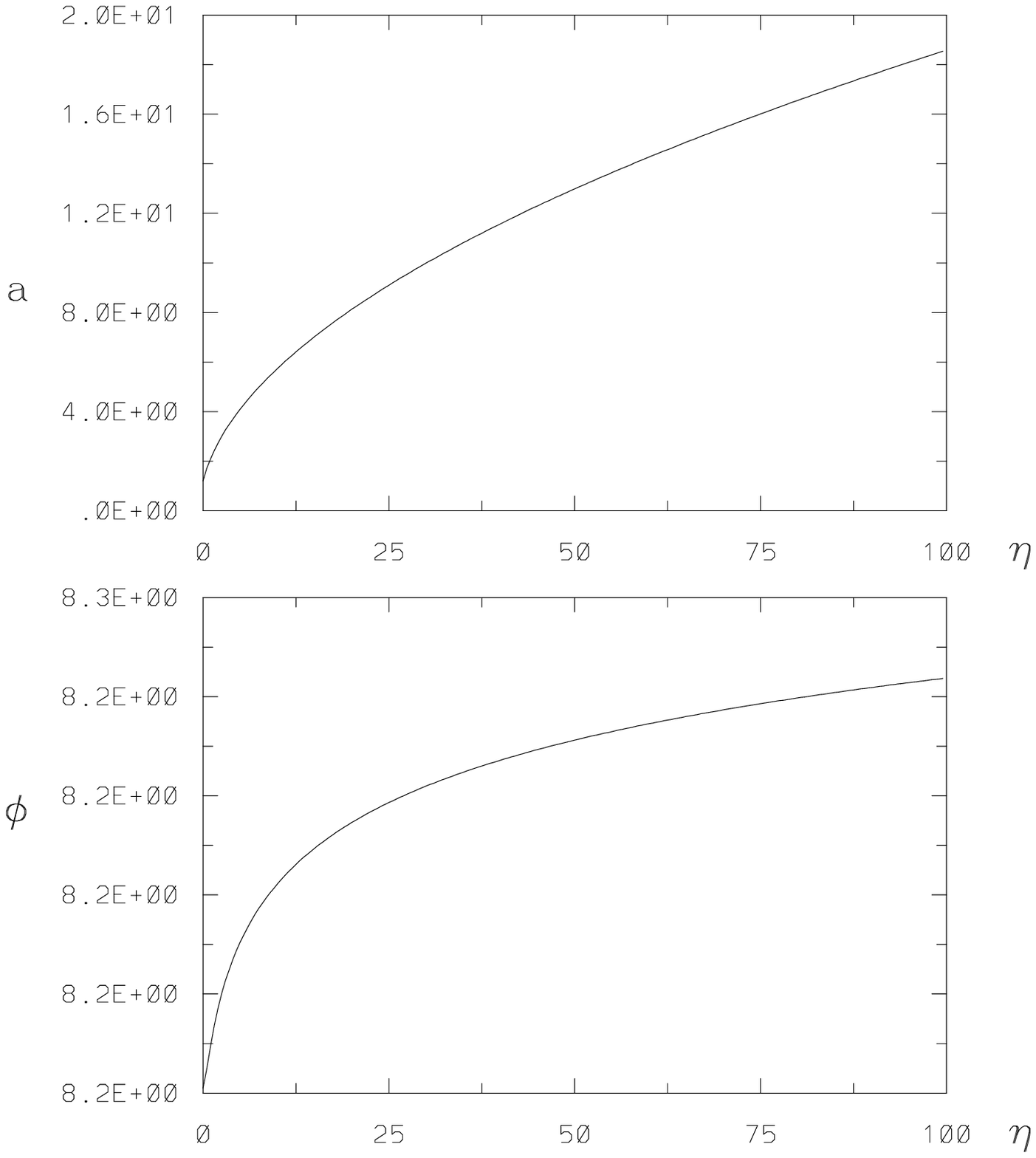}
\end{picture}
\vskip 1.5cm
\caption{The scale factor and BD field a function of time $\eta$.  The plots 
represent an open model with $\kappa_1=\kappa_2= -k =1$, and $\nu=\nu_3$ 
that implies $\omega=500$.
\label{fig3} }
\end{figure}

\begin{figure}
\unitlength 1in
\begin{picture}(6.5,6.5)(0,1)
\epsfxsize=5in \epsfbox{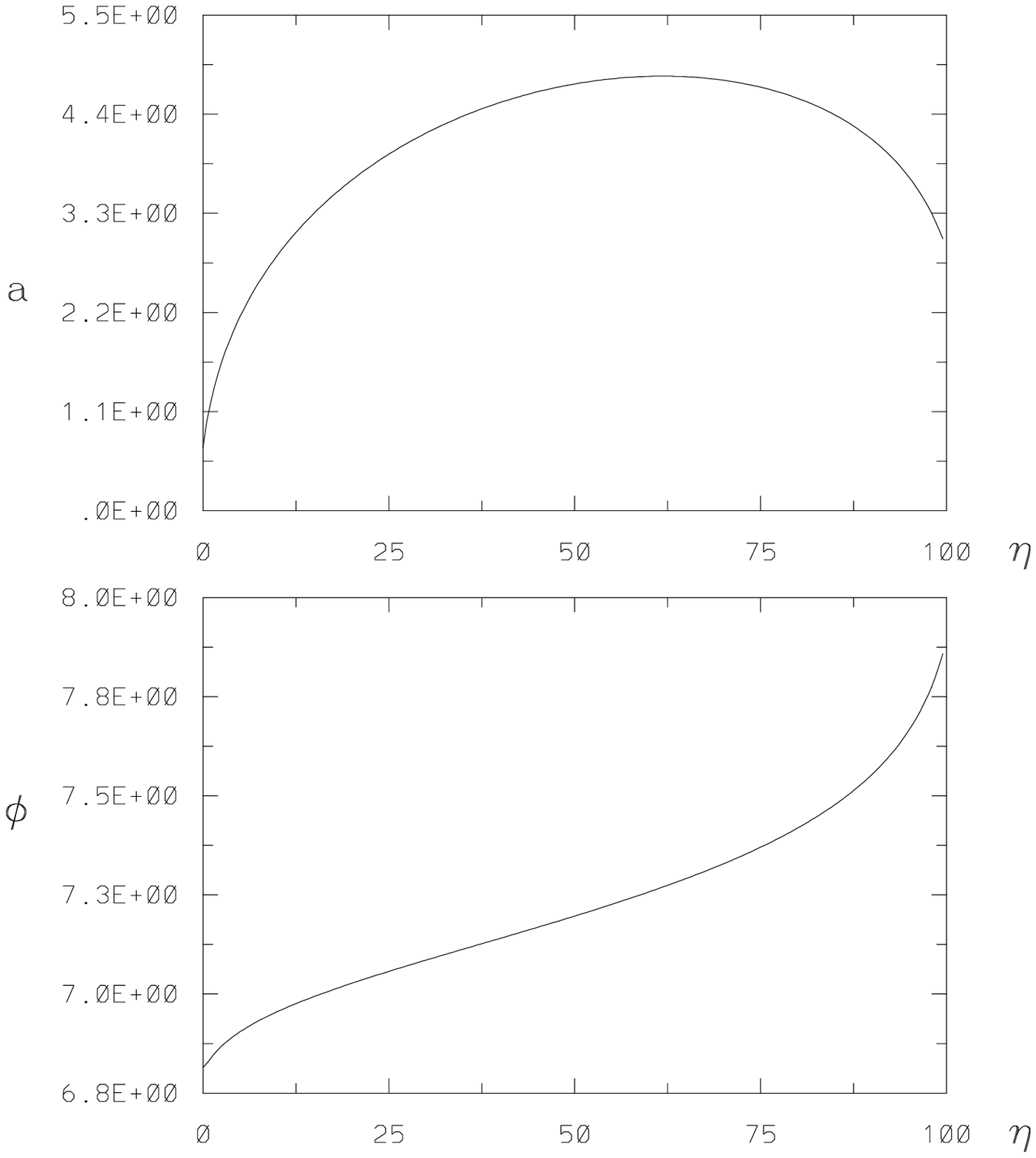}
\end{picture}
\vskip 1.5cm
\caption{The scale factor and BD field a function of time $\eta$.  The plots 
represent a closed model with $\kappa_1 = k =1$, $\kappa_2 = -0.8$, and 
$\nu=\nu_3$ that implies $\omega=500$.  The model eventually recollapses.
\label{fig4}}
\end{figure}

{}Figures \ref{fig3} and \ref{fig4} show an open and closed model, respectively,
with $\omega=500$. Again, there are two possible values of $\nu$, 
$\nu_3=-0.2988$ and $\nu_4=-0.3719$. In these figures we have chosen the value 
of $\nu_3$.   In figure \ref{fig3} we have chosen 
$\kappa_1=\kappa_2= -k =1$, whereas in figure \ref{fig4}  
$\kappa_1 = k =1$, $\kappa_2 = -0.8$.  The plots of figure \ref{fig3} 
do not follow Eq. (\ref{limg}) nor (\ref{lims}) since 
$2/(1-3\nu)$ is numerically very close to one, and the its asymptotic
behavior lies somewhere in between.

The relationship between the cosmic time $t$ and 
$\eta$, $d\eta \equiv a^{-3\nu} dt$, seems to be too complicated to be 
integrated in a closed form.  However, in the asymptotic limits given 
by Eqs. (\ref{limg}) and (\ref{lims}) one obtains that
\begin{eqnarray}
\label{teta}
t_{{\rm asimp1}} &=& \left(-k\right)^{\frac{3 \nu}{2(1-3\nu)}} \, 
(1-3\nu)^{\frac{1}{1-3\nu}}\, \eta^{\frac{1}{1-3\nu}} \\[2pt]
t_{{\rm asimp2}} &=& \frac{2 
\left[\frac{-2k}{\kappa_{2}}\right]^{\frac{3 \nu}{2(1-3\nu)}} 
(1-3\nu)^{2-\left(\frac{3\nu}{1-3\nu}\right)^{2}}
}{2-9\nu(1+\nu)}
\, \eta^{\frac{2-9\nu(1+\nu)}{2(1-3\nu)^{2}}} \, .
\end{eqnarray}
In both cases the functions are monotonic, and for 
$\nu_1$ and $\nu_3$ used in our plots, time grows as $\eta$ grows. Therefore,
our time parametrization is appropriate.   In this limit, one can express
our solutions in $\eta$ in terms of $t$ to get that:
\begin{eqnarray}
\label{solt1}
a_{{\rm asimp1}}    &=&    \sqrt{-k} \, t  \nonumber    \\[2pt]
\phi_{{\rm asimp1}} &=& m_{\nu} \, \left(\frac{1-3\nu}{1+3\nu}\right)^{2}
\kappa_2 \, \left(-k \right)^{- \frac{3(1-\nu)}{2(1-3\nu)}} 
\, = \, {\rm const.} 
\end{eqnarray}
and 
\begin{eqnarray}
\label{solt2}
a_{{\rm asimp2}} &=& 
\left[\frac{-2k}{\kappa_{2}}\right]^{\frac{1-3 \nu}{2-9\nu(1+\nu)}}
\left(1-3\nu\right)^{\frac{-2+3\nu(8-3\nu)(1-3\nu)}{2(1-3\nu)^{4}}} 
\left(1-9\nu(1+\nu)/2\right)^{\frac{1-9 \nu}{2-9\nu(1+\nu)}}
\, t^{\frac{1-9\nu}{2-9\nu(1+\nu)}} 
\nonumber\\[2pt]
\phi_{{\rm asimp2}} &=&   
\frac{2^{-\frac{(1+3\nu)(2+3\nu)}{2-9\nu(1+\nu)}}\, m_{\nu}}{(1+3\nu)^{2}}
\left[\frac{-\kappa_{2}}{k}\right]^{\frac{3(1+\nu)(1-3 \nu)}{2-9\nu(1+\nu)}}
\left(1-3\nu\right)^{\frac{4-21\nu-27 \nu^{2}}{2-9\nu(1+\nu)}}
\nonumber\\
&&\left(2-9\nu(1+\nu)\right)^{\frac{(1+3\nu)^{2}}{2-9\nu(1+\nu)}}
\, t^{\frac{(1+3\nu)^{2}}{2-9\nu(1+\nu)}} \, .
\end{eqnarray}
Eq. (\ref{solt1}) is a particular known flat solution in BD 
theory \cite{Lo84} or the $k=-1$ vacuum solution of GR. Eq. (\ref{solt2}) 
is the flat space, Nariai solution \cite{Na68} in which $\omega$ is given by
Eq. (\ref{wnu}).

\section{CONCLUSIONS \label{con}}

We have presented a set of differential equations written in rescaled 
variables that let us integrate a general equation to get Eq. (\ref{psi1}), 
valid for Bianchi type I, V and IX models, as well as for FRW models.  This 
equation is coupled to Eqs. (\ref{a123}) and (\ref{hi}), but for Bianchi 
type I and V models 
the anisotropic parameters ($h_i$'s) are constants (since their curvature 
terms are of FRW type) and  Eq. (\ref{psi1}) is decoupled.  This property 
allows one to find the general solution of Bianchi type I model and a 
particular solution of type V; both solutions are quadratic 
polynomials.  This solution let the models isotropize as time evolves, 
however, this can happens only for some  parameter ($\omega$, $\nu$) range.
The polynomial solution may also be valid for Bianchi type IX, but 
it is not proved yet.  If it were, isotropization would be also guaranteed.

We have found a new solution of Eq. (\ref{psi1}) valid for curved ($k \neq 0$) 
FRW cosmologies, that is, a solution with $h_i=0$.  This is
the general solution subject to the constraint given by Eq. (\ref{wnu}).  
Accordingly, we have analyzed two cases of physical interest: the 
case when $\omega=-1$, having some interest in string cosmology, that implies 
an equation of state of a quasi dust model ($\nu \,{}^{>}_{\sim} \, 0$), and the 
case when $\omega =500$, consistent with 
current BD local experimental constraints \cite{Wi93}, implying that 
$\nu \sim -1/3$.  Although in a different manner, both the scale factor and 
$\phi$ grow as a function of the parametrized time $\eta$ in all figures 
presented here.  One can find specific values of the 
constants $\nu, \kappa_2$ to have "closed" models without recollapse.  For 
this to happen $2/(1-3\nu)$ must be less than 1, i.e., $\nu$ must be 
negative.  This is a known effect of negative pressures.  

The new solution is non-inflationary and for
asymptotic times is of power-law type for the scale factor.  On the one hand, 
when the quantity $2/(1-3\nu)$ is greater than 1 the BD field 
tends to a constant value, then BD's dynamics evolves similar to GR's. 
One the other hand,  when $2/(1-3\nu)$ is smaller than 1, both scale factor
and BD field behave asymptotically with a power-law, and the solution is 
equal to Nariai solution for flat space.

Further solutions of Eq. (\ref{psi1}) are in order, which can be 
either within FRW cosmologies or Bianchi type V or IX models.  In particular,
the general solution with $h_i$'s constants should provide the general 
solution of both curved FRW cosmologies and Bianchi type V model.  

Finally, our results could be also of interest for 
$\omega(\phi)$-theories, where the value of the coupling parameter in some 
early cosmological era could have been rather different than its value 
nowadays, $\omega\ge 500$.


\end{document}